\documentclass[aps,pre,preprint,superscriptaddress,floatfix]{revtex4-1}
\usepackage{bm}
\usepackage{subfigure}
\usepackage{subfigure}
\usepackage{float}
\usepackage{amsmath,amsfonts,amssymb,graphicx}
\usepackage{graphicx}% Include figure files
\usepackage{dcolumn}% Align table columns on decimal point
\usepackage{color}
\begin{document}

\title{Asymmetry in interdependence makes a multilayer system more robust against cascading failures}

\author{Run-Ran Liu} \email{runranliu@163.com}
\affiliation{Alibaba Research Center for Complexity Sciences, Hangzhou Normal University, Hangzhou, Zhejiang 311121, China}

\author{Chun-Xiao Jia}
\affiliation{Alibaba Research Center for Complexity Sciences, Hangzhou Normal University, Hangzhou, Zhejiang 311121, China}

\author{Ying-Cheng Lai}
\affiliation{School of Electrical, Computer, and Energy Engineering, Arizona State University, Tempe, AZ 85287, USA}
\affiliation{Department of Physics, Arizona State University, Tempe, AZ 85287, USA}

\date{\today}

\begin{abstract}

Multilayer networked systems are ubiquitous in nature and engineering, and
the robustness of these systems against failures is of great interest. A main
line of theoretical pursuit has been percolation induced cascading failures,
where interdependence between network layers is conveniently and tacitly
assumed to be symmetric. In the real world, interdependent interactions are
generally asymmetric. To uncover and quantify the impact of asymmetry in
interdependence on network robustness, we focus on percolation dynamics
in double-layer systems and implement the following failure mechanism: once
a node in a network layer fails, the damage it can cause depends not only on
its position in the layer but also on the position of its counterpart
neighbor in the other layer. We find that the characteristics of
the percolation transition depend on the degree of asymmetry, where the
striking phenomenon of a switch in the nature of the phase transition
from first- to second-order arises. We derive a theory to calculate the
percolation transition points in both network layers, as well as the
transition switching point, with strong numerical support from synthetic
and empirical networks. Not only does our work shed light upon the factors
that determine the robustness of multilayer networks against cascading
failures, but it also provides a scenario by which the system can be designed
or controlled to reach a desirable level of resilience.

\end{abstract}

\date{\today}

\maketitle

\section{Introduction} \label{sec:intro}

It has been increasingly recognized that, because of the ubiquitous
presence of interdependence among different types of systems, a reasonable
understanding of a variety of complex phenomena in the real world requires
a description based on multilayer~\cite{KABGP:2014} or
interdependent~\cite{Havlin:2015,GSAJ:2012} networks.
Indeed, the functioning of a complex dynamical system, whether it be
physical, biological, or engineered, depends
not only on its own components, but also on other systems that are coupled
or interact with it~\cite{BPPSH:2010,Gao:2011,Gao:2012,LESL:2018}. Examples
of this sort abound in the real world, e.g., in social~\cite{Szell:2010},
technological~\cite{Ouyang:2014,Radicchi:2015}, and biological
systems~\cite{Klosik:2017}. In most existing models of multilayer networks,
the mutual interactions between a pair of network layers are treated as
symmetric. This assumption is ideal as the interactions between different
types of systems often are asymmetric. There then exists a gap between
current theoretical modeling/understanding and real world situations where
asymmetric interdependence is common. The purpose of this paper is to narrow
this gap by articulating a prototypical model of dynamics in
asymmetrically interacting multilayer networks and investigating its
robustness with the finding that interaction asymmetry can surprisingly
make the whole system significantly more robust.

To be concrete, we focus on a generic type of dynamical processes on multilayer
networked systems: cascading failures that attest most relevantly to the
robustness and resilience of the system. There is a large body of literature
on cascading failures in single layer complex networks~\cite{ML:2002,
Watts:2002,HK:2002,CLM:2004,ZPL:2004,ZPLY:2005,YCL:2005,HYY:2006,gleeson:2007,
HLC:2008,SBPBH:2008,YWLC:2009,WLA:2011,PBH:2011,HL:2011,Liu:2012,Gallos:2015,
KGAE:2018}, and there have also been efforts in cascading dynamics in
multilayer networks~\cite{PBH:2010,BPPSH:2010,Hu:2017PNAS,Shaw:2010,Zhao:2014,
NA:2015}. The unique feature that distinguishes cascading dynamics in multilayer
from those in single layer systems is that, in multilayer systems, failures can
propagate from one network layer to another and trigger large-scale failures
in an avalanche manner by the intricate strong node-to-node interaction
pattern across the network layers. Because of this, multilayer networks can
be vulnerable and collapse in an abrupt manner. While
protecting the hub nodes can be an effective strategy to mitigate cascading
failures in single layer networks, in interdependent systems this strategy is
less effective~\cite{Havlin:2015,GSAJ:2012}. Nonetheless, there are alternative
methods to generate robust interdependent networks even in the strong
dependence regime~\cite{Hu:2017PNAS,Shaw:2010,Zhao:2014,NA:2015,Liu_2019},
where robustness can be enhanced with second-order phase transitions through
mechanisms such as inter-similarity~\cite{Parshani:2011}, geometric
correlations~\cite{Kleineberg:2016,Kleineberg:2017}, correlated
community structures~\cite{Faqeeh:2018} and link
overlaps~\cite{Hu:2013,Cellai:2013}. In addition, it was found that the
vulnerability of interdependent networked systems can be reduced through
weakening the interlayer interaction~\cite{PBH:2010}.
It was also found that the topological properties of the network
layers composing a multilayer system, such as degree
correlations~\cite{Reis:2014,Zhou:2014,Valdez:2013,Min:2014},
clustering~\cite{Shao:2014,Huang:2013}, degree
distribution~\cite{Emmerich:2014,Yuan:2015}, inner-dependency~\cite{Wang:2018,
Liu:2016A}, and spatial embedding~\cite{Shekhtman:2014,Bashan:2013},
can affect the robustness of the whole multilayer system.
Another issue of great concern is how to destroy the largest mutually
connected component of a given multilayer network 
deliberately~\cite{Huang:2011,Dong:2012}. It was found that an 
effective way to destroy the giant component of a single network, i.e., 
destruction of the 2-core, does not carry over to multilayer networks. 
The methods of effective multiplex degree~\cite{Baxter2018}
and optimal percolation~\cite{Osat2017} were articulated for multilayer 
networks to find the minimal damage set that destroys the largest mutually 
connected component.

A tacit assumption employed in most previous models of cascading failures
in multilayer networks~\cite{PBH:2010,BPPSH:2010,Hu:2017PNAS,Shaw:2010,
Zhao:2014,NA:2015} is that the layer interdependence will cause a node to
fail completely should any of its neighboring nodes in the other layers
become non-functional. For convenience, we regard such interdependence as
``strong.'' As a result of the strong dependence, every pair of
interdependent nodes must be connected to the giant component in their
respective layer at the same time, motivating the introduction of the notion
of mutually connected components to characterizing the robustness of the whole
multilayer system~\cite{BPPSH:2010,Hackett:2016,KGAE:2018}. Analyses based
on the percolation theory~\cite{Broadbent:1957,Kirkpatrick:1973,Kesten:1982,
Stauffer:1992} revealed that the mutually connected component generically
undergoes a discontinuous phase transition as a function of the initial
random damage~\cite{BPPSH:2010,GSAJ:2012}. This result was somewhat
surprising because it is characteristically different from the continuous
percolation transition typically observed in single-layer
networks~\cite{AJB:2000,CEAH:2000,CEAH:2001,CNSW:2000}. Moreover, the
percolation theory provides a reasonable understanding of the catastrophic
cascading dynamics occurred in real interdependent infrastructure systems,
such as the interdependent system of power grids and telecommunication
networks~\cite{Rinaldi:2001}.

A deficiency of the assumption of ``strong'' interdependence is that nodes
across different layers in real world systems typically exhibit weaker types
of interdependence. For example, in a transportation system, passengers
can travel from city to city through a number of interdependent transportation
modes such as coaches, trains, airplanes, and ferries. When one mode becomes
unavailable, e.g., when the local airport is shut down, passenger flow into
the city may be decreased: some passengers destined for this city may cancel
their travel and the transferring passengers would switch to other cities to
reach their final destinations. Thus, although the disabling of the air
transportation route can have impacts on the function of the whole
interdependent networked system, transportation via other modes is still
available, i.e., total destruction will not occur and the system can still
maintain a certain level of functioning. For this particular example, the
interactions between the air travel network and other transportation network
layers are apparently asymmetric. Generally, many real-world infrastructure
systems such as electric power, water, or communication networks use backup
infrastructures and often have emergency management plans to survive losses
of interdependent services. In such a case, the failure of a node in one
layer can disable a number of links in other coupled layers, but not
necessarily cause the loss of {\em all} neighboring nodes and links. These
considerations motivated recent works on the consequences of ``weak''
interdependence in multilayer networked systems~\cite{LESL:2018,Liu:2016B}.
Another factor of consideration is that the impacts of a failed node on its
interdependent partners may depend not only on its position, but also on the
positions of the partners. That is, the strength of interdependence of two
nodes in different network layers is asymmetric, as in real infrastructure
systems. Intuitively, the origin of asymmetry in interdependence can be
argued, as follows. In a multilayer networked system, the ``important'' nodes
tend to be highly connected while the ``unimportant'' ones are less connected.
Probabilistically, the failure of an ``unimportant'' node thus would not
have a great impact on the ``important'' nodes, but the failure of an
``important'' node is more likely to have significant effects on the
``unimportant'' nodes. To investigate the consequence of asymmetrically
interdependent interactions may thus lead to a better understanding of
robustness and resilience of multilayer networks in the real world.

In this paper, we articulate a class of percolation dynamical models for
multilayer networks incorporating asymmetric interdependence and investigate
the effects of the asymmetry on the robustness of the whole system. In our
model, the strength of the interdependence of nodes with different degrees
on its partners in different layers is not identical to the strength in the
opposite direction. We introduce a generic parameter $\theta$ to characterize
the degree or extent of asymmetry of two interdependent network layers.
Intuitively, it may occur that asymmetry can make the system more vulnerable
to catastrophic failures. For example, disabling some nodes that exert more
influence on its partner nodes than the other way around is more likely to
lead to failures of these nodes, making cascading failures more probable.
However, counter intuitively, we find that increasing the degree of asymmetry
can dramatically improve the robustness of the whole multilayer system.
In particular, when the highly connected nodes in one layer depend less
on the nodes of lower degrees in the other layer than the dependence in
the opposite direction, the robustness of the system can be improved
significantly as compared with the counterpart system with perfectly symmetric
interdependence~\cite{BPPSH:2010}. Quantitatively, as the degree
of asymmetry is systematically increased, the system undergoes a remarkable
switch from a first-order percolation transition to a second-order one. We
develop an analytic theory to predict the characteristic changes in the nature
of the phase transition as induced by asymmetry and the transition points,
with strong numerical support based on percolation dynamics in both synthetic
and empirical networks. Our results suggest that, in order to enhance network
robustness and resilience (as in designing a multilayer infrastructure system),
introducing an appropriate level asymmetric interaction among the
interdependent layers can be advantageous.

\section{Model} \label{sec:model}

We consider a percolation process on an interdependent system with two network
layers, denoted as $A$ and $B$, each having the same number $N$ of nodes. The
functioning of node $a_{i}$ $(i=1,\cdots,N)$ in network $A$ depends on the
functioning of the counterpart node $b_{i}$ in network $B$, and vice versa.
When $a_{i}$ fails, each link of its dependent partner $b_{i}$ will be
maintained or disabled with probability $\alpha^{b}_{i}$ or $1-\alpha^{b}_{i}$,
respectively. Similarly, if $b_{i}$ in $B$ fails, the links of its dependent
partner $a_{i}$ in network $A$ will be intact or disabled with probability
$\alpha^{a}_{i}$ or $1-\alpha^{a}_{i}$, respectively. The probabilities
$\alpha^{a}_{i}$ and $\alpha^{b}_{i}$ thus characterize the interdependence
strength of node $a_{i}$ on $b_{i}$ and vice versa, respectively. When
the values of $\alpha^{a}_{i}$ or $\alpha^{b}_{i}$ approach one, the
interdependence between the two nodes is the weakest, where failures are unable
to spread from one network layer to another. The opposite limit where the
values of $\alpha^{a}_{i}$ and/or $\alpha^{b}_{i}$ approach zero corresponds
to the case of the strongest possible interdependence. In general, the values
of $\alpha^{a}_{i}$ and $\alpha^{b}_{i}$ are different and degree dependent.
One way to define these parameters is
\begin{equation} \label{beta}
\alpha^{a}_{i} = \frac{(k^{a}_{i})^{\theta}}{(k^{a}_{i})^{\theta}+(k^{b}_{i})^{\theta}} \ \ \mbox{and} \ \
\alpha^{b}_{i} = \frac{(k^{b}_{i})^{\theta}}{(k^{a}_{i})^{\theta}+(k^{b}_{i})^{\theta}}.
\end{equation}
where $k^{a}_{i}$ and $k^{b}_{i}$ are the degrees of nodes $a_{i}$ and
$b_{i}$, respectively, and $\theta$ is a parameter that controls the asymmetry
of the interdependent interactions. In particular, for $\theta=0$, the
interdependence between nodes $a_{i}$ and $b_{i}$ is symmetric:
$\alpha^{a}_{i}=\alpha^{b}_{i}$, regardless of the nodal degrees.
For $\theta>0$, the interdependence is weak of a high degree node in one
network layer on a low degree node in the other network layer and the
interdependence of a low degree node in one layer on a high degree node in
the other layer is strong. The opposite situation occurs for $\theta < 0$,
where the interdependence of a high (low) degree node in one layer on a
low (high) degree node in the counter layer is strong (weak).
As described in Introduction, in real multilayer networks, the failure of
a less connected node would not have a great impact on the well connected
nodes, but the failure of a well connected node is more likely to have
significant effects on the less connected nodes. This means the case of
negative $\theta$ values may seldom appear and positive $\theta$ values are
more general in realistic scenarios. Tuning the value of $\theta$ enables a
systematic analysis of the effects of asymmetry in interdependence on the
robustness of the whole multilayer system.

We start the percolation process by randomly removing a fraction $(1-p)$ of
the nodes of networks $A$ and $B$ independently. In each network layer, the
links connected to the removed nodes are simultaneously removed. This is
the case where an initial attack occurs in the two network layers
simultaneously. The removal of nodes in one network will cause some nodes
to be isolated from the giant component and to fail, and the failure can spread
across the whole system through an iterative process. In each iteration,
disconnecting certain nodes from the giant component of, e.g., network $A$,
will cause some nodes to be isolated from the giant component of network
$B$ through the destruction of some of their links, which in turn will
induce more link destruction and nodal failures in $A$. When the process of
failure stops, the whole system reaches a stable steady state. The sizes
$S^{A}$ and $S^{B}$ of the giant components in the final state of the network
layers $A$ or $B$ can be used to measure the robustness of the whole
system~\cite{BPPSH:2010,LESL:2018}.

\section{Theory} \label{sec:theory}

We develop a theory to understand the asymmetry-induced switch between
first- and second-order phase transitions and to predict the transition
points. Let $p_{k}^{A}$ and $p_{k}^{B}$ be the degree distributions of
network layers $A$ and $B$, respectively, where the average degrees are
given by $\langle k\rangle^{A} = \sum_{k}p_{k}^{A}k$ and
$\langle k\rangle^{B} = \sum_{k}p_{k}^{B}k$.
The final sizes $S^{A}$ and $S^{B}$ of the respective giant components in
layers $A$ and $B$ in the steady state can be solved by using a self-consistent
probabilistic approach. In particular, define $R^{A}$ ($R^{B}$) to be the
probability that a randomly chosen link in network $A$ ($B$) belongs to its
giant component. Suppose we randomly choose a node $a_{i}$ of degree
$k^{a}_{i}$ in network $A$. The probability of functioning of this node
depends the state of its interdependent neighbor $b_{i}$ in network $B$.
If $b_{i}$ is disabled, each of its links can be maintained with the
probability $ \alpha^{a}_{i}$, so the probability that this link leads to
the giant component in $A$ is $\alpha^{a}_{i}R^{A}$. The viable probability
of node $a_{i}$ in $A$ is thus given by
$p[1-(1-\alpha^{a}_{i} R^{A})^{k^{a}_{i}}]$
if its interdependent neighbor $b_{i}$ is not viable. If $b_{i}$ is functional,
the viable probability of node $a_{i}$ is $p[1-(1- R^{A})^{k^{a}_{i}}]$.
With the quantity $R^{B}$, we can get the viable probability of node $b_{i}$
as $p[1-(1- R^{B})^{k^{b}_{i}}]$. Taking into account the probability
distributions of $k^{a}_{i}$ and $k^{b}_{i}$, the viable probability of a
random node in $A$ is
\begin{eqnarray} \label{eq:SA}
	S^{A} & = & p^{2}[1-\sum_{k^{a}_{i}}p_{k^{a}_{i}}^{A}(1-R^{A})^{k^{a}_{i}}] [1-\sum_{k^{b}_{i}}p_{k^{b}_{i}}^{B}(1- R^{B})^{k^{b}_{i}}] \\ \nonumber
	& + &
 p\sum_{k^{a}_{i}}\sum_{k^{b}_{i}}p^A_{k^{a}_{i}}p_{k^{b}_{i}}^{B}[1-(1-\alpha^{a}_{i} R^{A})^{k^{a}_{i}}]\{1-p[1-(1- R^{B})^{k^{b}_{i}}]\},
\end{eqnarray}
where the first and second terms denote the cases where $b_{i}$ is viable
and not viable, respectively. Similarly, the viable probability of a node
$b_{i}$ in network $B$ is
\begin{eqnarray} \label{eq:SB}
S^{B} & = & p^{2}[1-\sum_{k^{b}_{i}}p_{k^{b}_{i}}^{B}(1-R^{B})^{k^{b}_{i}}] [1-\sum_{k^{a}_{i}}p_{k^{a}_{i}}^{A}(1- R^{A})^{k^{a}_{i}}] \\ \nonumber
& + & p\sum_{k^{a}_{i}}\sum_{k^{b}_{i}}p^A_{k^{a}_{i}}p_{k^{b}_{i}}^{B}[1-(1-\alpha^{b}_{i} R^{B})^{k^{b}_{i}}]\{1-p[1-(1- R^{A})^{k^{a}_{i}}]\}.
\end{eqnarray}
Following a randomly chosen link in network $A$, we can arrive at a node
$a_{j}$ with degree $k^{a}_{j}$. If its interdependent neighbor $b_{j}$ of
degree $k^{b}_{j}$ in network $B$ is viable, the random link can lead to
the giant component with the probability $p[1-(1-R^{A})^{k^{a}_{j}-1}]$. If
$b_{j}$ is not viable, each link of node $a_{j}$ is reserved with the
probability $\alpha^{a}_{j}$, and the random link can lead to the giant
component with the probability
$p\alpha^{a}_{j}[1-(1-\alpha^{a}_{j} R^{A})^{k^{a}_{j}-1}]$. These
considerations lead to the following self-consistent equation for $R^{A}$:
\begin{eqnarray} \label{eq:RA}
	R^{A} & = & p^{2}[1-\sum_{k^{a}_{j}}\frac{p_{k^{a}_{j}}^{A}k^{a}_{j}}{\langle k\rangle^{A}}(1-R^{A})^{k^{a}_{j}-1}] [1-\sum_{k^{b}_{j}}p_{k^{b}_{j}}^{B}(1- R^{B})^{k^{b}_{j}}] \\ \nonumber
	& + &
  p\sum_{k^{a}_{j}}\sum_{k^{b}_{j}}\frac{p_{k^{a}_{j}}^{A}k^{a}_{j}}{\langle k\rangle^{A}}p_{k^{b}_{j}}^{B}\alpha^{a}_{j}[1-(1-\alpha^{a}_{j} R^{A})^{k^{a}_{j}-1}]\{1-p[1-(1- R^{B})^{k^{b}_{j}}]\}.
\end{eqnarray}
A self-consistent equation for $R^{B}$ can be obtained in a similar way. We get
\begin{eqnarray} \label{eq:RB}
	R^{B} & = & p^{2}[1-\sum_{k^{b}_{j}}\frac{p_{k^{b}_{j}}^{B}k^{b}_{j}}{\langle k\rangle^{B}}(1-R^{B})^{k^{b}_{j}-1}] [1-\sum_{k^{a}_{j}}p_{k^{a}_{j}}^{A}(1- R^{A})^{k^{a}_{j}}] \\ \nonumber
	& + &
  p\sum_{k^{a}_{j}}\sum_{k^{b}_{j}}p_{k^{a}_{j}}^{A}\frac{p_{k^{b}_{j}}^{B}k^{b}_{j}}{\langle k\rangle^{B}}\alpha^{b}_{j}[1-(1-\alpha^{b}_{j} R^{B})^{k^{b}_{j}-1}]\{1-p[1-(1- R^{A})^{k^{a}_{j}}]\}.
\end{eqnarray}

\begin{figure}
\centering
\includegraphics[width=\linewidth]{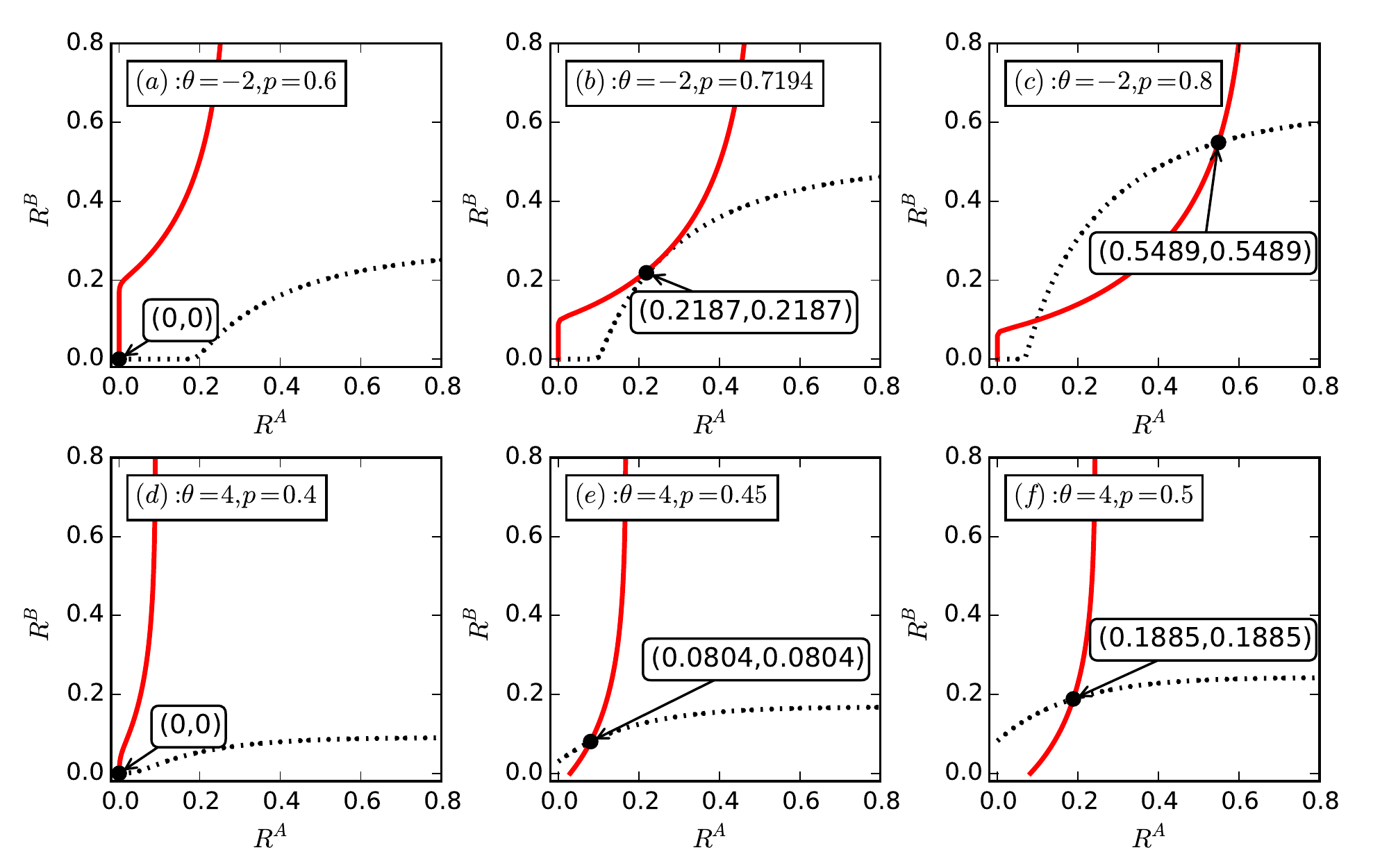}
\caption{ {\em Solutions of the self-consistent equations for the probabilities
that a random node belongs to the giant component in a double layer system}.
Shown are the graphical solutions of Eqs.~(\ref{eq:RA}) and (\ref{eq:RB}) for
different values of $\theta$ and $p$, as marked by the black dots. 
(a-c) Results for $p=0.6$, $p=0.7194$ and $p=0.8$, respectively, for 
$\theta = -2$, and (d-f) the solutions for $p=0.4$, $p=0.45$, and $p=0.5$, 
respectively, for $\theta = 4$. The average degree is $\langle k\rangle = 4$.}
\label{fig:GS}
\end{figure}

Figure~\ref{fig:GS} shows, for random networks $A$ and $B$ with a Poisson
degree distribution~\cite{Bollobas:1985, Molloy:1995}
$p_k=e^{-\langle k\rangle}\langle k\rangle^k/k!$, graphical solutions of
$R^{A}$ and $R^{B}$ for different values of $\theta$ and $p$. For simplicity,
we consider the case where $A$ and $B$ have the identical degree distribution:
$p_{k}^{A}=p_{k}^{B}\equiv p_{k}$. For $\theta=-2$, there is a trivial
solution at the point $(R^{A}=0,R^{B}=0)$ for $p=0.6$, indicating that
both networks $A$ and $B$ are completely fragmented. For $p=0.7194$,
the solutions are given by the tangent point $(0.2187,0.2187)$, giving rise
to a discontinuous change in both $R^{A}$ and $R^{B}$ that is characteristic
of a first-order percolation transition. For $\theta=4$, the crossing point
for $R^{A}$ and $R^{B}$ changes continuously from $(0,0)$ to some nontrivial
values, indicating a continuous (second order) percolation transition.

The critical point for both first- and second-order types of percolation
transition can be obtained, as follows. For $p_{k}^{A}=p_{k}^{B}\equiv p_{k}$,
we have $\langle k\rangle^{A}=\langle k\rangle^{B}\equiv \langle k\rangle$,
$R^{A}=R^{B}\equiv R$. Equation~(\ref{eq:RA}) or (\ref{eq:RB}) can then
be reduced to
\begin{eqnarray} \label{eq:RAB}
	R & = & p^{2}[1-\sum_{k^{a}_{j}}\frac{p_{k^{a}_{j}}k^{a}_{j}}{\langle k\rangle}(1-R)^{k^{a}_{j}-1}] [1-\sum_{k^{b}_{j}}p_{k^{b}_{j}}(1- R)^{k^{b}_{j}}]
	\\ \nonumber
	& + &
  p\sum_{k^{a}_{j}}\sum_{k^{b}_{j}}\frac{p_{k^{a}_{j}}k^{a}_{j}}{\langle k\rangle}p_{k^{b}_{j}}\alpha^{a}_{j}[1-(1-\alpha^{a}_{j} R)^{k^{a}_{j}-1}]\{1-p[1-(1- R)^{k^{b}_{j}}]\}\equiv h(R).
\end{eqnarray}
For the first-order transition, the straight line $y = x$ and the curve
$R = h(R)$ from Eq.~(\ref{eq:RAB}) become tangent to each other at the point
$(R_{c},R_{c})$, at which the derivatives of both sides of Eq.~(\ref{eq:RAB})
with respect to $R$ are equal:
\begin{equation} \label{eq:C1}
\frac{dh(R)}{dR}|_{R=R_{c},p=p_{c}^{I}}=1.
\end{equation}
Equations~(\ref{eq:RAB}) and (\ref{eq:C1}) can be solved numerically to
yield the first-order percolation transition point $p^{I}_{c}$.

In the regime of second-order percolation transition, the probability $R$
tends to zero as $p$ approaches the percolation point $p^{II}_{c}$. We can
use the Taylor expansion of Eq.~(\ref{eq:RAB}) for $R \equiv \epsilon\ll 1$:
\begin{equation} \label{eq:TAB}
h(\epsilon)= h'(0)\epsilon+\frac{1}{2}h''(0)\epsilon^{2}+O(\epsilon^{3})=\epsilon.
\end{equation}
Since $\epsilon \in (0,1)$, we obtain
$h'(0)+\frac{1}{2}h''(0)\epsilon+O(\epsilon^{2})=1$ by dividing both sides
of Eq.~(\ref{eq:TAB}) by $\epsilon$. Neglecting high order terms of
$\epsilon$, we have that the second-order percolation point $p^{II}_{c}$
is determined by the solutions of
\begin{equation} \label{eq:TC}
h'(0)=p_{c}^{II}\sum_{k^{a}_{j}}\sum_{k^{b}_{j}}\frac{p_{k^{a}_{j}}k^{a}_{j}}{\langle k\rangle}p_{k^{b}_{j}}(\alpha^{a}_{j})^{2}(k^{a}_{j}-1)=1.
\end{equation}

If, for any node, we have $\alpha_{j}^{a} \rightarrow 1$, there will be no
interdependence across the network and Eq.~(\ref{eq:TC}) can be reduced to the
case of a single-layer network. In this case, the percolation transition point
becomes $p^{II}_c=\langle k\rangle /\langle k(k-1)\rangle$, which is the same
result for single-layer generalized random networks and can be validated to
be consistent with the previous ones~\cite{CEAH:2000,CEAH:2001}. Since the
interdependence strength of a pair of nodes across two network layers is
determined by their degrees, 
the situation $\alpha_{j}^{a} \rightarrow 0.5$ arises if all pairs of 
interdependent nodes have exactly the same degree, which provides a special 
symmetrical case for any given value of $\theta$.

When the conditions for the first- and second-order transitions are
satisfied simultaneously, i.e., $p^{I}_{c} = p^{II}_{c}$, the percolation
transition switches from first to second order (or vice versa). Substituting
$p^{II}_{c}$ from Eq.~(\ref{eq:TC}) into Eq.~(\ref{eq:TAB}), we have
\begin{equation} \label{eq:C2}
\frac{1}{2}h''(0)\epsilon^{2}+O(\epsilon^{3})=0.
\end{equation}
For the first-order percolation transition, $\epsilon_{c}$ is always nontrivial
and Eq.~(\ref{eq:C2}) is not applicable any more. Apparently, if the system
undergoes a second-order percolation transition, the value of $\epsilon$ is at
the transition point $\epsilon_{c}=0$ and Eq.~(\ref{eq:C2}) is naturally
satisfied. On the boundary between the first- and second-order percolation
transitions, the value of $\epsilon_{c}$ is negligibly small. We have
\begin{eqnarray} \label{eq:C3}
	h''(0) & = & p^{II}_{c}[2p^{II}_{c}\sum_{k^{a}_{j}}p_{k^{a}_{j}}k^{a}_{j}(k^{a}_{j}-1)
    +p_{c}^{II}\sum_{k^{a}_{j}}\sum_{k^{b}_{j}}\frac{p_{k^{a}_{j}}k^{a}_{j}}{\langle k\rangle}p_{k^{b}_{j}}(\alpha^{a}_{j})^{2} \\ \nonumber
	& - & \sum_{k^{a}_{j}}\sum_{k^{b}_{j}}\frac{p_{k^{a}_{j}}k^{a}_{j}}{\langle k\rangle}p_{k^{b}_{j}}(k^{a}_{j}-1)(k^{a}_{j}-2)(\alpha^{a}_{j})^{3}
- p_{c}^{II}\sum_{k^{a}_{j}}\sum_{k^{b}_{j}}\frac{p_{k^{a}_{j}}k^{a}_{j}}{\langle k\rangle}p_{k^{b}_{j}}(k^{a}_{j}-1)(\alpha^{a}_{j})^{2}]=0.
\end{eqnarray}
For a given degree distribution $p_{k}$, we can obtain the crossover point
$\theta_{c}$ of the first- and second-order transition points by solving
Eq.~(\ref{eq:C3}) numerically.

\section{Results} \label{sec:results}

\begin{figure}
\centering
\includegraphics[width=\linewidth]{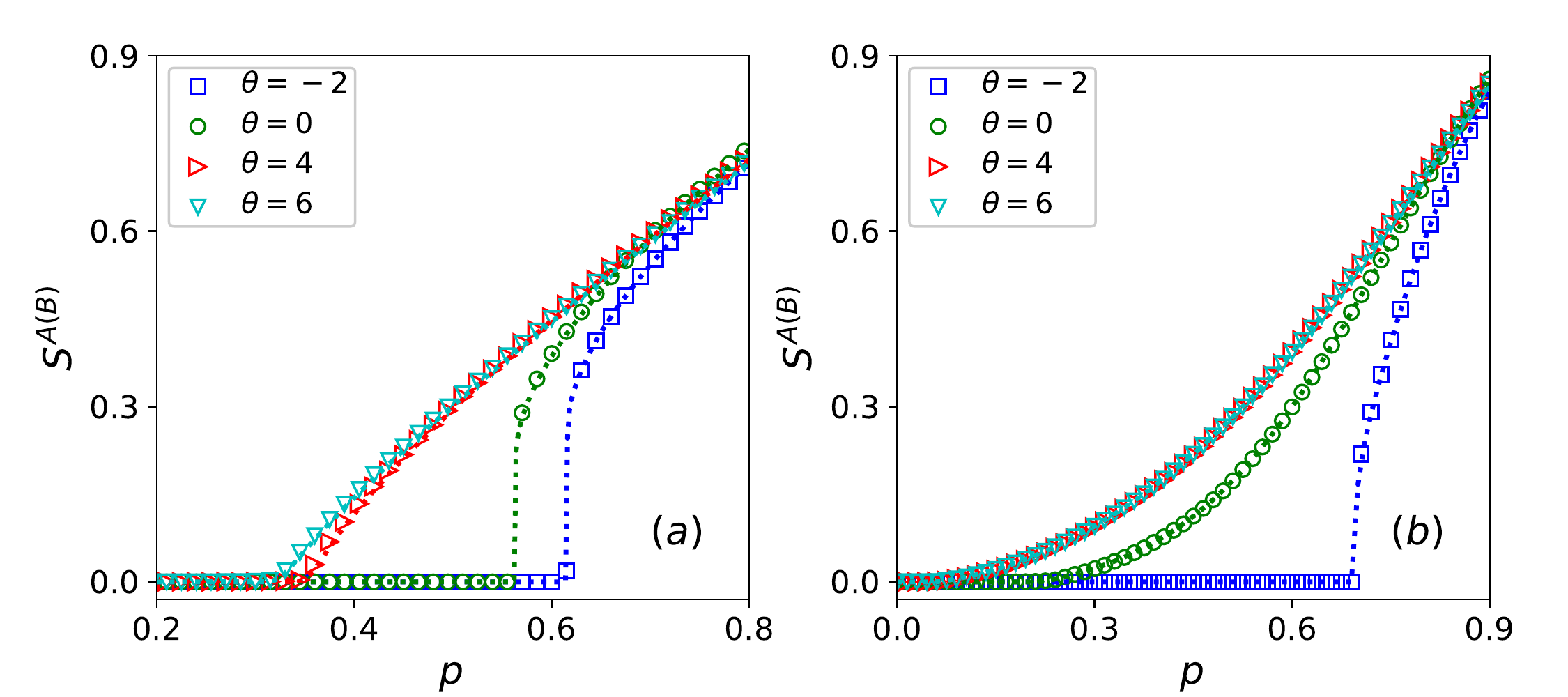}
\caption{ {\em Simulation results for first- and second-order percolation
transitions on interdependent random 
(a) and scale-free (b) networks}. Shown are the
fractions $S^{A}$ and $S^{B}$ of nodes in the respective giant component at
the end of a cascading process as a function of $p$ for $\theta=-2,0,4,6$. The
results are obtained by averaging over $40$ independent realizations, where
the network size is $N=5\times10^{5}$ with the average degree
$\langle k \rangle = 5$ for both random and scale-free networks. For
scale-free interdependent networks, the minimum degree is $2$ and the
power-law exponent of degree distribution is $-2.3$. The dotted curves
underlying the symbols represent the theoretical predictions obtained from
Eq.~(\ref{eq:RAB}), all agreeing well with the numerical results.}
\label{fig:Gvsp}
\end{figure}

Figures~\ref{fig:Gvsp}(a) and \ref{fig:Gvsp}(b) show the sizes of the giant
components in network layers $A$ and $B$, denoted by $S^{A}$ and $S^{B}$,
versus the fraction $p$ of initially preserved nodes for interdependent
random and scale-free networks, respectively. For a negative value of the
asymmetry parameter $\theta$ (e.g., $\theta = -2$), $S^{A}$ and $S^{B}$
percolate discontinuously at a threshold $p_{c}^{I}$. For a positive value
of $\theta$ (e.g., $\theta = 4$), the networks $A$ and $B$ percolate
continuously with a reduced value of the transition point $p_{c}^{II}$,
leading to a crossover in the percolation transition and a higher degree
of system robustness. Further increase in the value of $\theta$ leads to
little change in the transition point $p_{c}^{II}$, indicating that
the ability to enhance the system robustness by increasing the value of
$\theta$ is saturated, for both interdependent random and scale-free
networks. Theoretical predictions are also included in Fig.~\ref{fig:Gvsp},
which agree with the numerical results quite well.

\begin{figure}
\centering
\includegraphics[width=\linewidth]{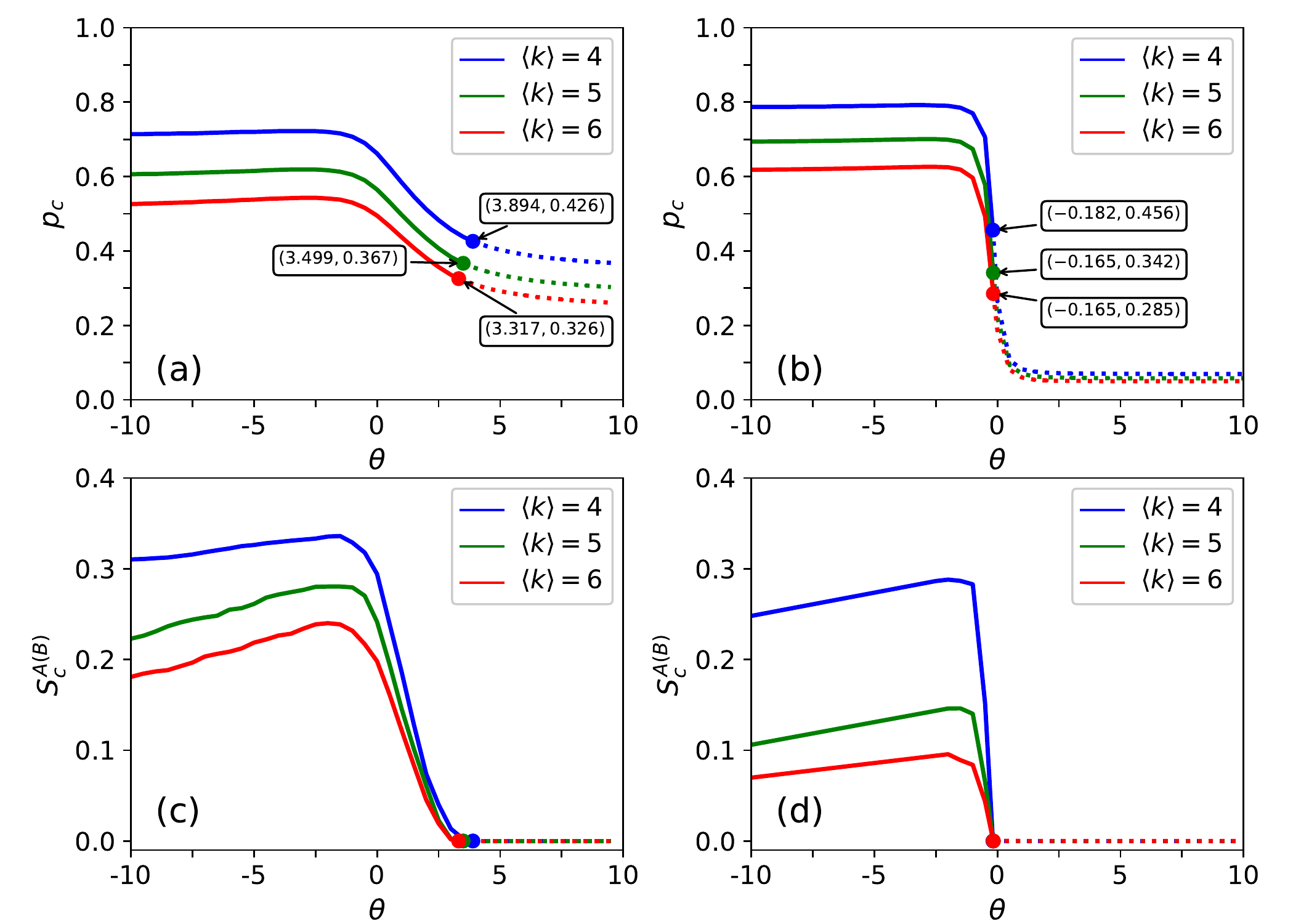}
\caption{ {\em Dependence of the percolation transition point and the critical
size of giant component at the percolation transition point on the asymmetry
parameter $\theta$}.
(a,b) The transition point $p_{c}$ versus $\theta$ for a system of 
interdependent random and scale-free networks, respectively. 
(c,d) The critical size $S^{A(B)}_{c}$ of the giant component at the 
percolation transition point versus $\theta$ for a system of interdependent
random and scale-free networks, respectively. For both random and scale-free 
networks, the average degree $\langle k\rangle$ is $4$, $5$ and $6$ (For 
scale-free networks, the minimum degree is $2$ and the corresponding 
power-law exponent of degree distribution is $-2.6$, $-2.3$, or $-2.1$.) 
For each $\langle k\rangle$ value, there exists a critical point $\theta_{c}$, 
marked by the large solid dots, that divides the $\theta$ interval into two 
subregions with distinct types of phase transitions: first-order
(solid curves) and second-order (dotted curves), respectively. At first-order
phase transition points, the critical size $S^{A(B)}_{c}$ of the giant 
component is nonzero and the transition is abrupt. At the second-order phase 
transition points, the critical size $S^{A(B)}_{c}$ of the giant component 
is zero and the transition is continuous.
Increasing the value of the asymmetry
parameter $\theta$ from a negative to some positive value has two advantages:
(i) a decreased value of the critical transition point $p_c$ regardless of
the nature of the transition (i.e., first or second order), indicating that
more nodes can be removed before the occurrence of a phase transition, and
(ii) a switch in the transition from first to second order, where the former
is often catastrophic while the latter can be benign.}
\label{fig:transition}
\end{figure}

Figures~\ref{fig:transition}(a) and \ref{fig:transition}(b) show the
percolation transition points $p_{c}^{I}$ $(p_{c}^{II})$ versus $\theta$ for
a random and a scale-free interdependent networked system, respectively.
In Fig.~\ref{fig:transition}(a), for each average-degree value tested, the
phase diagram is divided into two distinct regions by a critical point:
for $\theta < \theta_{c}$, the transition is discontinuous (first-order)
while it is continuous (second-order) for $\theta>\theta_{c}$ with
relatively smaller values of the percolation threshold $p_{c}^{II}$. Similar
behaviors occur for scale-free interdependent networks, as shown in
Fig.~\ref{fig:transition}(b). For both types of interdependent networks,
as $\theta$ is increased, the percolation transition point $p_{c}^{I}$
$(p_{c}^{II})$ moves towards lower values, indicating that more nodes
can be removed before a phase transition occurs and, consequently,
the whole system becomes more robust. Nonetheless, as $\theta$ is further
increased, the transition point $p_{c}^{I}$ $(p_{c}^{II})$ becomes saturated.
A distinct feature in the changes of $p_{c}^{I}$ and $p_{c}^{II}$ versus
$\theta$ is that, near the crossover point $\theta_{c}$, the transition
point for the scale-free networked system is more sensitive to asymmetry
in the interdependent interactions than the random networked system. This
result suggests that, near $\theta_{c}$, the robustness of the scale-free
interdependent network can be compromised by the asymmetry.
Figures~\ref{fig:transition}(c) and \ref{fig:transition}(d)
show the the critical size $S^{A(B)}_{c}$ of giant component at the
percolation transition point as functions of the asymmetrical parameter
$\theta$. Above the switch point $\theta_{c}$, $S^{A(B)}_{c}$ is finite
characterizing a discontinuous phase transition, whereas below the switch
point $\theta_{c}$, $S^{A(B)}_{c}$ is zero and the system percolates
as a continuous phase transition.

For lower values of the asymmetrical parameter $\theta$, large-degree
nodes in one network depend on the small-degree nodes in the
other network with a large coupling strength. In this case, the failure of
a low-degree node in one network can destroy a high-degree node in the other
network. The small-degree nodes are sensitive to nodal or link removal
and have a high risk to fail in a cascading process. Although the large-degree
nodes are ``stubborn'', they are more destructive than the low-degree nodes
in case of failures. That is, a low value of $\theta$ can reinforce the
dependence of the destructive nodes on high-risk nodes, amplifying the
systematic risk for the whole interdependent system. As the value of $\theta$
is increased to become positive, the dependence strength of the high-degree
nodes on the low-degree nodes is reduced, making the system relatively more
robust.

Since the interdependence of a pair of nodes is controlled by
the degree difference of them in terms of the asymmetrical parameter $\theta$,
we introduce a parameter $\omega$ to control the fraction of
overlapping links and the degree difference of interdependent nodes in
a double-layer network. An overlapping link is defined in terms of a pair of
links that connect two pairs of interdependent nodes in different network
layers, respectively. In particular, say there exist a link that connects
two nodes $a_{i}$ and $a_{j}$ in layer $A$. The link connecting nodes $b_{i}$
and $b_{j}$ in layer $B$ is an overlapping link. For
$\omega \rightarrow 0$, there is no overlapping link and the system reduces
to the one studied above. For $\omega \rightarrow 1$, all the links
are overlapping links and the degrees of nodes in network $A$ are the same
as the degrees of their respective interdependent nodes.
In this case, the interdependence is symmetric and $\alpha=0.5$ 
for any given value of $\theta$.

\begin{figure}
\centering
\includegraphics[width=\linewidth]{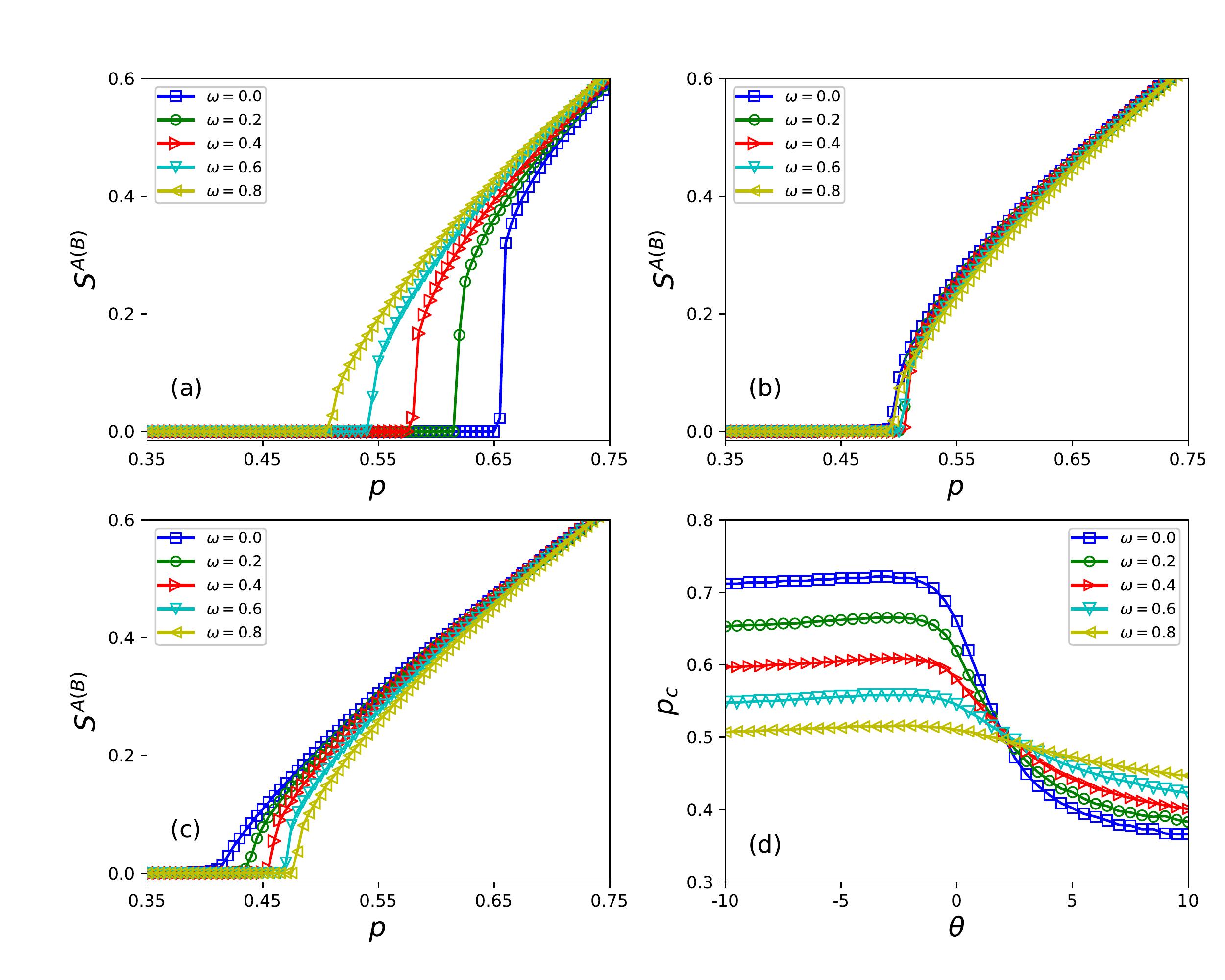}
\caption{ {\em Simulation results of percolation transitions
in random networks with overlapping links}.
(a-c) The fractions $S^{A}$ and $S^{B}$ of nodes in the respective giant
component at the end of a cascading process as a function of $p$ for
different fractions of overlapping links for $\theta=0$, $\theta=2$
and $\theta=4$, respectively. (d) The percolation transition point $p_{c}$ 
as a function of $\theta$ for different values of $\omega$.
The results are obtained by averaging over $40$ statistical realizations.
The network size is $N=5\times10^{5}$ and the average degree is 
$\langle k \rangle = 4$.}
\label{fig:overlap}
\end{figure}

Figure~\ref{fig:overlap} shows the simulation results for percolation
transitions in random networks with overlapping links. We find that, for
a fixed value of $\omega$, the value of the percolation transition point
$p_c$ decreases with the increase in the asymmetrical parameter $\theta$.
This means that the system is robust when large-degree nodes in one
network layer depend strongly on large-degree nodes in the other layer,
but the system becomes vulnerable when there is strong interdependence between
large-degree nodes in one layer and small-degree nodes in the other
layer. We also find that the curves of the percolation transition point $p_c$
versus $\theta$ for different values of $\omega$ intersect at the point
$\theta_c \approx 2$, as shown in Fig.~\ref{fig:overlap}(d). For
$\theta < \theta_c$, the percolation point $p_c$ decreases with the increase
in the value of $\omega$, indicating that an increase in the fraction of
overlapping links makes the system more robust. However, for $\theta >\theta_c$,
the percolation point $p_c$ increases and the system becomes less robust
as the value of $\omega$ is increased.
As the networks become fully overlapped ($\omega\rightarrow 1$), the
differences in the degrees of the interdependent nodes across the network
layers decrease and the value of the interdependence strength approaches 
$0.5$ irrespective of the value of $\theta$. In this case, increasing $\theta$
will not lead to any appreciable change in the asymmetry of the interdependence
strength among the network layers. This is the reason why the overlap does 
not contribute to enhancing the system robustness in the region of large 
$\theta$ values, as exemplified in Fig.~\ref{fig:overlap}.
These results suggest that the role of overlapping links in the robustness
of a system depends on the value of the asymmetrical parameter $\theta$.

\begin{figure}
\centering
\includegraphics[width=\linewidth]{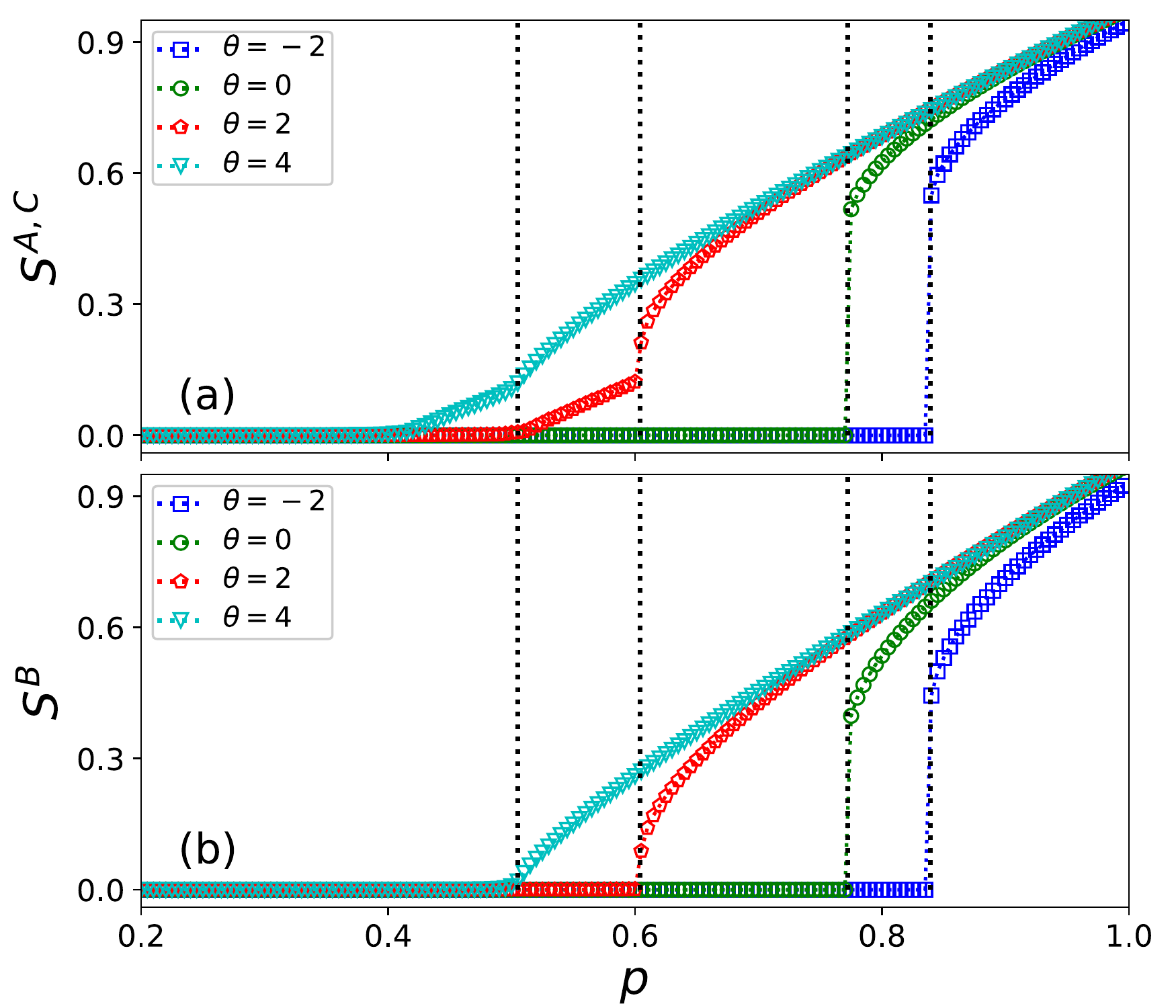}
\caption{{\em Percolation transitions in three-layer random
networks}. The fractions $S^{A(C)}$ (a) and $S^{B}$ (b) of nodes, respectively,
in the corresponding giant component at the end of a cascading process as a
function of $p$ for different values of the asymmetrical parameter $\theta$.
The straight vertical dotted lines denote the positions of percolation
transition of $B$. The results are obtained by averaging over 40 statistical
realizations. The network size is $N=5\times10^{5}$ and the average degree
is $\langle k \rangle = 4$.}
\label{fig:three-layer}
\end{figure}

What about percolation transitions in multilayer systems with more than
two layers? To address this question, we study three-layer systems with
asymmetrical interdependencies. To be concrete, we consider the following
configuration of interdependence among the three layers ($A$, $B$ and $C$):
layer $A$ depends on layer $B$, layer $B$ depends on layer $C$, but layers
$A$ and $C$ have no direct dependence on each other. Depending on the
extent of asymmetrical interdependencies, multiple percolation transitions
can occur. Figure~\ref{fig:three-layer} shows that the transition point
$p_c$ decreases and the system becomes more robust as the value of the
asymmetrical parameter $\theta$ is increased. We also find that multiple
percolation transitions occur for relatively large values of $\theta$, and
layers $A$ and $C$ percolate first, followed by layer $B$, after which another
phase transition occurs in layers $A$ and $C$. However, for small values of
$\theta$, the phenomenon of multiple percolation transitions disappears
because the three layers tend to percolate at the same point. These results
are consistent with those in Ref.~\cite{LESL:2018}, demonstrating
that both asymmetry in the interdependence and layer position can be
important for the functioning of the multilayer interdependent systems.

A practical implication is that the asymmetrical parameter can be
exploited for modulating or controlling the characteristics of the percolation
transition~\cite{LESL:2018}. In particular, for relatively low degree of
asymmetry (e.g., $\theta = -2$), the percolation transitions are abrupt
and discontinuous. In this case, the interdependent system is not resilient
and is likely to collapse suddenly as random nodal failures or intentional
attacks intensify. To improve the resilience of the system, a larger value
of $\theta$ can be chosen (e.g., $\theta = 4$) to make the system collapse,
if at all inevitable, to occur in a continuous fashion. While the whole system
still collapses eventually, the manner by which the collapse occurs is benign
and the value of the critical point $p_{c}^{II}$ is smaller as compared
with that associated with first-order phase transitions.

\begin{figure}
\centering
\includegraphics[width=\linewidth]{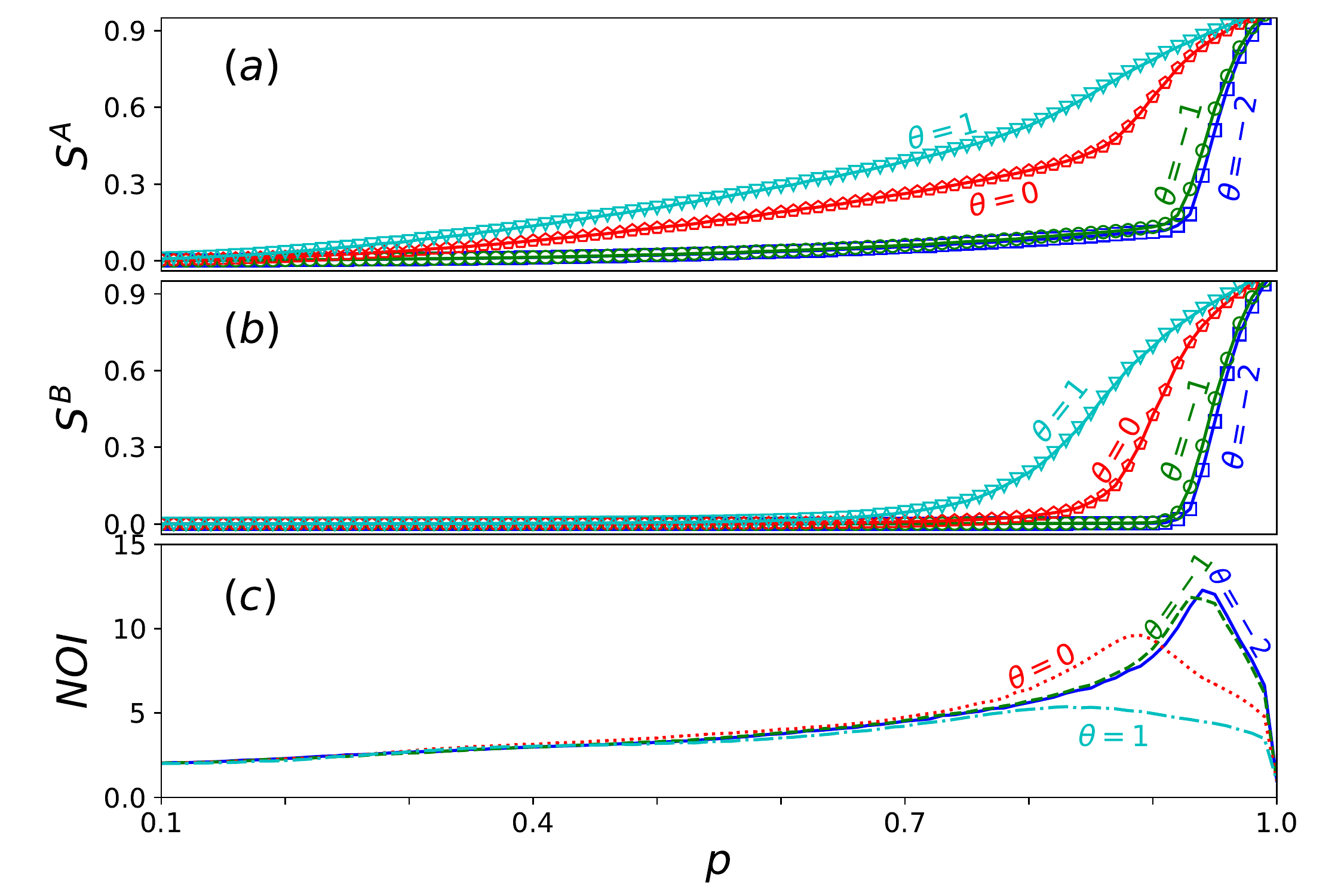}
\caption{ {\em Role of asymmetric interdependence in enhancing the robustness
of an Internet-power system in a controllable manner}.
(a,b) The sizes of the giant components of the autonomous 
system of the Internet and the power grid versus $p$ for different values of 
$\theta$, respectively. (c) The number of iterations (NOI) as a functions of 
$p$ for different values of $\theta$ for the Internet and power grid networks.
The data points are the result of averaging over 1000 statistical 
realizations.}
\label{fig:robust}
\end{figure}

We demonstrate the role of asymmetric interdependence in enhancing the
robustness of multilayer networked systems in a controllable manner by
studying a real world networked system with asymmetrical interdependence:
an autonomous systems of the Internet and the power grid of the Western
States of USA (data sets available at http://konect.uni-koblenz.de/).
The autonomous systems of the Internet consists of $6474$ nodes~\cite{AS}
and the power grid has $4941$ nodes~\cite{natureSW1998}
with each being either a generator, a transformer or a substation. We
randomly choose a number of nodes from the power grid as the dependent
partners of the nodes in the autonomous level Internet, and define an
interdependence link between a power grid node and an Internet node until all
the selected power grid nodes and the Internet nodes are connected. The
dependency strengths of the power grid and the Internet nodes are assigned
according to Eq.~(\ref{beta}). That is, if a node in the power grid fails,
the Internet node that depends on it will suffer a loss of some links
because of the interdependence and reserve some links because of the
existing buffering effect, and vice versa. Figure~\ref{fig:robust} shows
the sizes of the giant components of the Internet and the power grid
versus $p$ for different values of $\theta$. For negative values
of $\theta$ (e.g., $\theta=-2$), the sizes of the giant
components reduce drastically as $p$ is decreased from one, as indicated by a
relative large number of iterations in the cascading process. While for a
relatively large positive value of $\theta$, the changes in the sizes of the
giant components versus $p$ are smooth, signified by fewer iterations in
the cascading process.

\begin{figure}
\centering
\includegraphics[width=\linewidth]{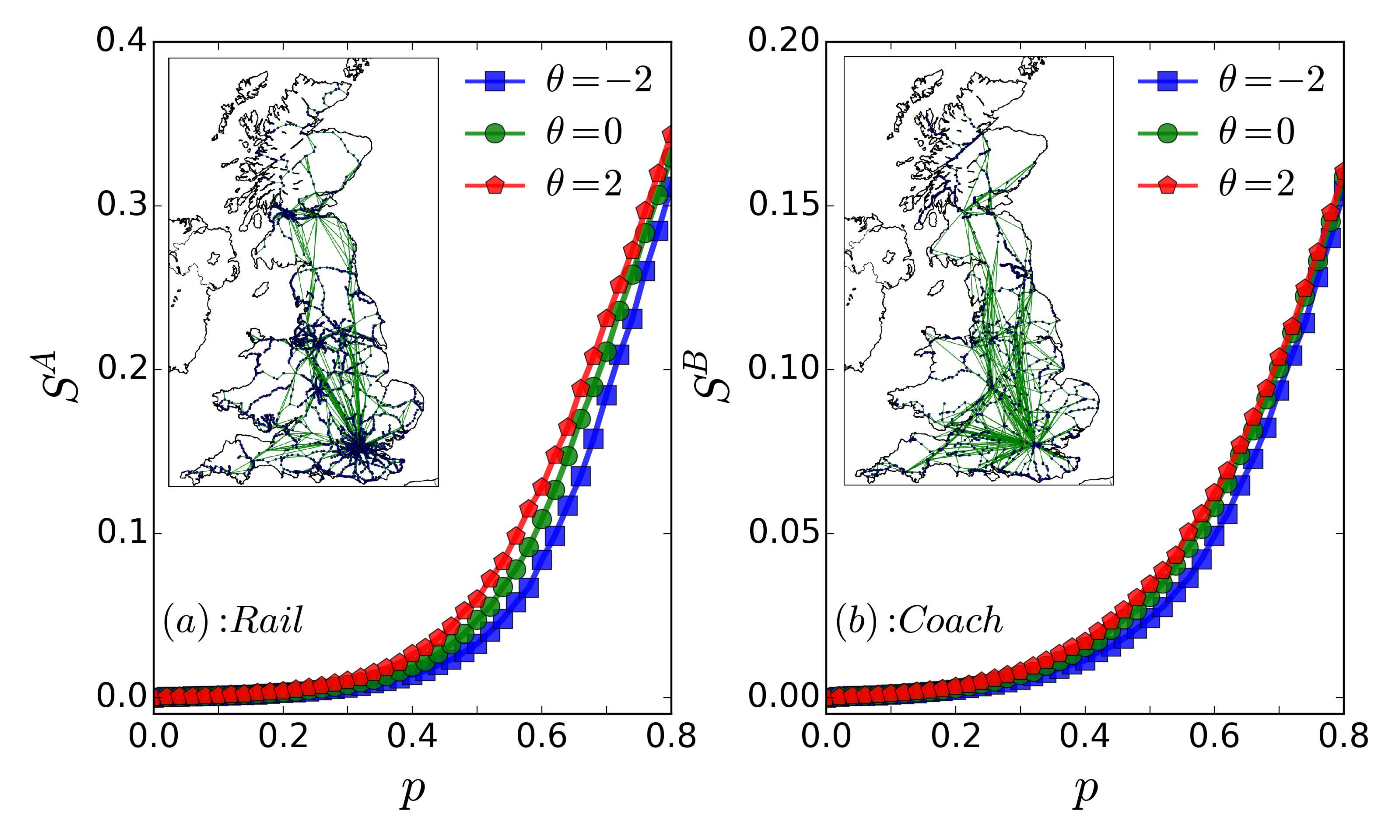}
\caption{{\em Role of asymmetric interdependence in enhancing the robustness
of a transportation system in a controllable manner in rail and coach
transportation system in Great Britain}. Shown are the sizes
of the giant components of the rail layer (a) and the coach
layer (b) versus $p$ for different values of $\theta$,
respectively. The data points are the result of averaging over 1000
statistical realizations. In each panel, the upper left
inset shows the actual route map for the corresponding layer.}
\label{fig:rail-coach}
\end{figure}

Another example is the rail and coach transportation system in
Great Britain~\cite{Gallotti:2015}, which includes rail, coach, ferry
and air transportation layers. Here we use the data in October 2011 to set up
the multilayer network and conduct our numerical experiments. An
analysis of the data shows that the rail and coach occupy $98.1\%$ of all the
inter-urban connections, and ferry and air transportation account for the
remaining $1.9\%$. We thus focus on the former to construct a two-layer
network. Due to the interdependence of passenger flows between different
traffic layers, the dependence strengths of the coach station and the rail
station are assigned according to Eq.~(\ref{beta}). Figure~\ref{fig:rail-coach}
shows the sizes of the giant components of the rail and coach layers
versus $p$ for different values of $\theta$. For negative values of $\theta$
(e.g., $\theta=-2$), the sizes of the giant components are always lower than
that of the case when $\theta$ is positive (e.g., $\theta=0$ or $\theta=2$).

These results demonstrate that the proposed principle of asymmetrical
interdependence can be effective for enhancing the robustness of the
functioning of the interdependent system. Especially, an appropriate
amount of asymmetry in the interdependence can be quite beneficial to
preventing a first-order transition from happening which leads to sudden,
system-wide failures.

\section{Discussion} \label{sec:discussion}

Interdependent multilayer networked systems in the real world are
generally asymmetric in layer-to-layer interactions, and the asymmetry
will inevitably have an impact on the robustness of the whole system. In
most existing works on dynamical processes in multilayer networks, the mutual
interactions between a pair of network layers are treated as symmetric,
giving rise to a knowledge gap in our general understanding of the dynamics on
multilayer networked systems and their robustness against failures and/or
attacks. The present work aims to narrow the gap by investigating a generic
type of dynamics on multilayer networks: cascading failures.

For simplicity and to facilitate analysis but without sacrificing generality,
we have studied double-layer networked systems and focus on the percolation
dynamics, where a cascading process can be triggered by random removal of
nodes and links, which can cause a dramatic reduction in the sizes of the giant
components in both network layers and possibly lead to total fragmentation
of the system. There are two characteristically distinct failure scenarios: as
the fraction of removed nodes is increased, the sizes of the giant components
will inevitably reduce to near zero values, either in a discontinuous or in a
continuous manner, corresponding to a first- or a second-order phase
transition, respectively. From the standpoint of network robustness and
resilience, a first-order transition is undesired as the system can become
fragmented abruptly. Even if a total system breakdown is inevitable, it is
desired that the process occurs gradually and continuously, which is
characteristic of second-order phase transitions.

Our main finding is that asymmetric interdependence can shift the critical
point of the percolation induced phase transition in a desirable way and,
strikingly, can affect the nature of the transition. In particular, as the
degree of asymmetry is systematically tuned, the system can undergo a switch
from a first-order percolation transition to a second-order one. Qualitatively,
this can be understood, as follows. When the nodes with large degrees in one
network layer depend highly on the nodes of small degrees in the other
network, the failure of a low-degree node in the latter can destroy a high
degree node in the former. In this case, the whole system can be quite fragile
due to the relative abundance of the small degree nodes, where a first-order
phase transition is expected. Quite the contrary, when nodes of large degrees
in one network layer depend on the large degree nodes in the other layer,
a second-order percolation transition arises and the system is robust. We
have developed a theory to predict the phase transition points and provide
strong numerical support with synthetic network models. To demonstrate the
relevance of our work and finding to the real world, we have also studied
the double-layer system of Internet and power grid with randomly assigned
one-to-one interdependence and the rail-coach transportation system.

From the point of view of design and control, our finding implies that the
degree of asymmetry (or symmetry) of interdependence can be exploited to
enhance the robustness of multilayer networks against cascading failures.
This can be especially meaningful in engineering design of complex
infrastructure systems that are intrinsically multilayer structured, or in
biological systems where the interdependent interaction strength may be
tuned biochemically.

\section*{Acknowledgment}

RRL was supported by the National Natural Science Foundation of China
under Grant Nos.~61773148 and 61673150. YCL would like to acknowledge
support from the Vannevar Bush Faculty Fellowship program sponsored by
the Basic Research Office of the Assistant Secretary of Defense for
Research and Engineering and funded by the Office of Naval Research
through Grant No.~N00014-16-1-2828.

%\bibliography{Percolation1}

 \newcommand{\noop}[1]{}

\end{document}